\documentclass[%
 reprint,
superscriptaddress,
 amsmath,amssymb,longbibliography,
 aps,
]{revtex4-2}

\usepackage{graphicx}
\usepackage{dcolumn}
\usepackage{bm}
\usepackage{siunitx}
\usepackage{xcolor}
\usepackage{braket}
\usepackage{siunitx}

\begin{document}

\preprint{APS/123-QED}

\title{Measuring the visual angle of polarization-related entoptic phenomena using structured light}

\author{C. Kapahi}
\email{c3kapahi@uwaterloo.ca}
\affiliation{Institute for Quantum Computing, University of Waterloo,  Waterloo, ON, Canada, N2L3G1}
\affiliation{Department of Physics, University of Waterloo, Waterloo, ON, Canada, N2L3G1}
\author{A. E. Silva} 
\affiliation{School of Optometry and Vision Science, University of Waterloo, Waterloo, ON, Canada, N2L3G1}
\author{D. G. Cory}
\affiliation{Institute for Quantum Computing, University of Waterloo,  Waterloo, ON, Canada, N2L3G1}
\affiliation{Department of Chemistry, University of Waterloo, Waterloo, ON, Canada, N2L3G1}
\author{M. Kulmaganbetov}
\affiliation{Centre for Eye and Vision Research, Hong Kong}
\author{M. Mungalsingh}
\affiliation{School of Optometry and Vision Science, University of Waterloo, Waterloo, ON, Canada, N2L3G1}
\author{D. A. Pushin}
\email{dmitry.pushin@uwaterloo.ca}
\affiliation{Institute for Quantum Computing, University of Waterloo,  Waterloo, ON, Canada, N2L3G1}
\affiliation{Department of Physics, University of Waterloo, Waterloo, ON, Canada, N2L3G1}
\affiliation{Centre for Eye and Vision Research, Hong Kong}
\author{T. Singh}
\affiliation{Centre for Eye and Vision Research, Hong Kong}
\author{B. Thompson}
\affiliation{School of Optometry and Vision Science, University of Waterloo, Waterloo, ON, Canada, N2L3G1}
\affiliation{Centre for Eye and Vision Research, Hong Kong}
\author{D. Sarenac}
\email{dsarenac@uwaterloo.ca}
\affiliation{Institute for Quantum Computing, University of Waterloo,  Waterloo, ON, Canada, N2L3G1}
\affiliation{Centre for Eye and Vision Research, Hong Kong}

\date{\today}

\begin{abstract}
    The ability to perceive polarization-related entoptic phenomena arises from the dichroism of macular pigments held in Henle's fiber layer of the retina and can be inhibited by retinal diseases such as age-related macular degeneration, which alter the structure of the macula.
    Structured light tools enable the direct probing of macular pigment density through the perception of polarization-dependent entoptic patterns.
    Here, we directly measure the visual angle of an entoptic pattern created through the illumination of the retina with a structured state of light and a perception task that is insensitive to corneal birefringence.
    The central region of the structured light stimuli was obstructed, with the size of the obstruction varying according to a psychophysical staircase.
    The perceived size of the entoptic pattern was observed to vary between participants, with an average visual angle threshold radius of  $9.5^\circ \pm 0.9^\circ$, 95\% C.I. = [$5.8^\circ$, $13^\circ$], in a sample of healthy participants.
    These results (with eleven azimuthal fringes) differ markedly from previous estimates of the Haidinger's brush phenomenon's extant (two azimuthal fringes), of $3.75^\circ$, suggesting that higher azimuthal fringe density increases pattern visibility.
    The increase in apparent size and clarity of entoptic phenomenon produced by the presented structured light stimuli may possess greater potential to detect the early signs of macular disease over perception tasks using uniform polarization stimuli.
\end{abstract}

\maketitle


\section{Introduction}

The human perception of polarized light has long been an area of active research, dating back to the identification of the Haidinger's Brush phenomenon in 1844~\cite{haidinger1846beobachtung}, continuing with the Maxwell and Helmholtz models of human polarization perception~\cite{maxwell1850manuscript, von2013treatise}, to modern investigations, motivated by the observation that human polarization perception is inhibited in patients with early-stage age-related macular degeneration (AMD)~\cite{wang2022mathematical, Wang2022MathematicalSensitivity, misson2020polarization}.
Despite this research interest, current clinical diagnosis of AMD does not utilize polarization perception, in part because existing polarization perception diagnostic tests show significant false-positive AMD detection, that is, healthy participants are not always able to perceive these polarization patterns~\cite{o2021seeing, tseng2021emergence}.
AMD remains the leading cause of blindness in people over the age of 60 in the developed world~\cite{mottes2022haidinger}, and a significant number of people over 60 show symptoms of AMD that are not detected in standard vision care~\cite{neely2017prevalence}.

Over the past five decades, the term ``structured light'' has evolved from referring to the generation of intensity patterns at a particular camera plane by diffractive optics to a wide variety of techniques for creating non-trivial phase patterns in coherent beams of light and other non-optical waves~\cite{rubinsztein2016roadmap, bliokh2023roadmap, chen2021engineering}.
Recently, the field of structured light has become associated with orbital angular momentum (OAM) states which possess an azimuthal phase-ramp from $0$ to some multiple (also called the OAM number) of $2\pi$.
The development of numerous components capable of generating the phase vortex indicative of OAM states has led to applications in information transmission, matter manipulating, and material science~\cite{BarnettBabikerPadgett, mair2001entanglement, wang2012terabit, friese1996optical, brullot2016resolving, simpson1997mechanical, sarenac2018generation, schwarz2020talbot}.
Furthermore, advancements in the generation of structured states of light have bridged the gap between structured light research and biological material characterization, including the study of biological tissue~\cite{biton2021oam}, and the study of human polarization perception~\cite{sarenac2020direct, sarenac2022human, gassab2023conditions, kominis2022quantum}.

Structured light tools enable the generation of polarization states which elicit entoptic profiles inaccessible with traditional forms of light~\cite{sarenac2020direct, sarenac2022human}.
Here, we present the first direct measurement of the visual angle of entoptic patterns through a psychophysical perception task coupled with retinal imaging of structured light profiles.
The extent of the entoptic pattern generated by polarization-coupled OAM states with OAM value $\ell = 13$ was measured by obstructing the central region of the structured light state, with a variable radius obstruction.
A forced-choice 2-up/1-down staircase was used to vary the size of the obstruction to determine the apparent radius of the entoptic pattern.
This radius could then be converted to apparent size in degrees of visual angle with reference to retinal images taken with an imaging device using structured light illumination.
By projecting the polarization-coupled state directly onto the retina, we remove the propagation effects seen in ~\cite{sarenac2020direct} and enable the use of variable obstructions.
Furthermore, we enable the generation of entoptic patterns selectively on particular regions of the retina.
Leveraging these advantages, we report that the perceived entoptic pattern generated by an $\ell=13$ OAM state is $9.5^\circ \pm 0.9^\circ$ in diameter, while Haidinger's Brush extends $3.75^\circ$~\cite{coren1971use}.
This increase in apparent entoptic pattern size and clarity suggests that structured states may provide significantly increased utility as a diagnostic tool over polarization perception methods utilizing Haidinger's Brush.

\section{Theory}

\begin{figure*}
    \centering
    \includegraphics[width=0.95
    \textwidth]{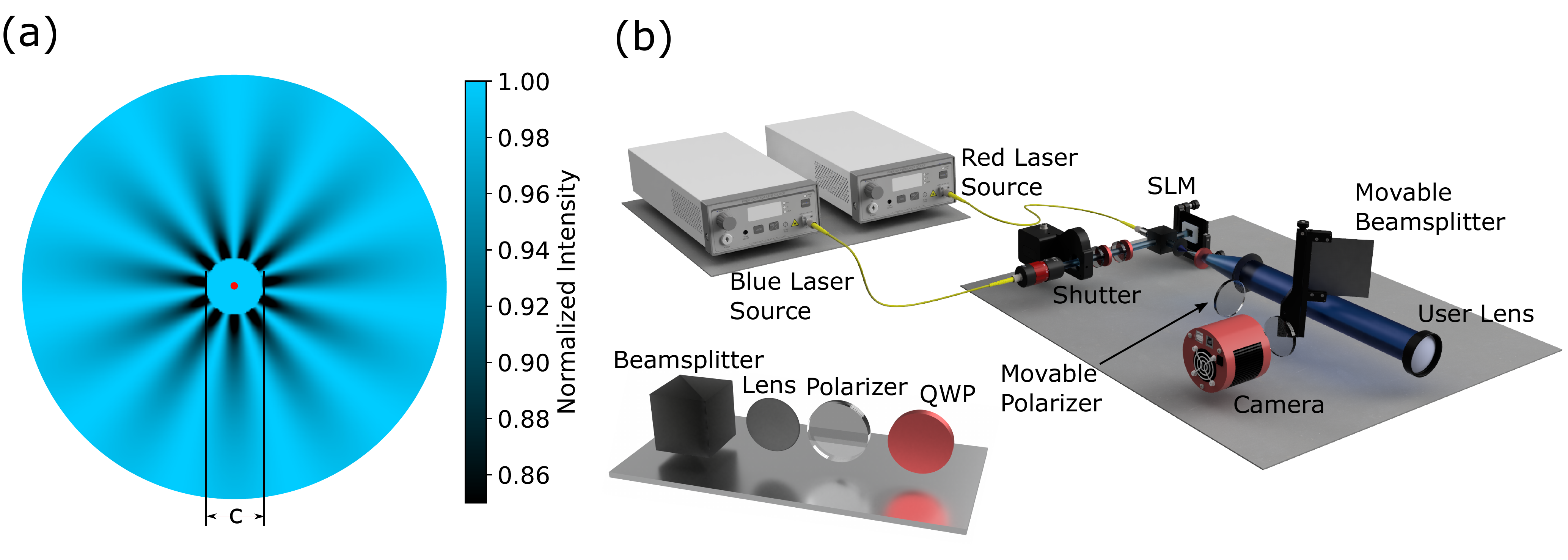}
    \caption{
    (a) Simulation of the entoptic pattern generated by a $\ell=13$ spin-coupled OAM state, as shown in Equation~\ref{eqn:spin-orbit-state}.
    The intensity of this entoptic pattern was generated using the macular pigment optical density (MPOD) model shown in Equation~\ref{eqn:mpod}.
    The state viewed by participants has a red guide light at the center of the orbital angular momentum state.
    The size of the central circular obstruction is shown here as $c$.
    (b) Monochromatic light from a \SI{450}{\nano\meter} diode laser is prepared in a circularly polarized state, followed by a polarizer and quarter waveplate, each mounted on separate rotation stages.
    Following this, the beam is prepared in a polarization state which is orthogonal to the optic axis of the spatial light modulator (SLM).
    Light can then be prepared in the polarization-coupled OAM state described in Equation~\ref{eqn:spin-orbit-state}.
    A two-lens imaging system is used to expand the beam for viewing through a Volk 20D Binocular Indirect Ophthalmoscopy lens.
    Retinal imaging is enabled by the translation stage mounted plate beamsplitter placed before the Volk lens.
    When placed in the beam, this plate beamsplitter directs reflected light from the retina through a linear polarizer to the camera-lens imaging system.
    The structured light imaging microscope enables the measurement of structured light pattern dimensions directly on a participant's retina.
    }
    \label{fig:apparatus}
\end{figure*}

While several physiological mechanisms through which humans perceive polarized light have been proposed~\cite{mcgregor2014human}, recent consensus is that axons located in Henle's fiber layer of the retina cause this effect~\cite{temple2015perceiving, misson2015human}.
These axons contain pigment molecules, such as lutein, with an average orientation radial to the central axis of the fiber~\cite{temple2019haidinger}.
Within the macula, these axons lie parallel to the surface of the retina and are oriented radially~\cite{lujan2011revealing}.
The dichroic properties of lutein lead to Henle's fiber layer acting as a radially oriented dichroic optical element in the region of the retina where these fibers lie parallel to the retinal surface~\cite{bone1984macular}.
A model for the macular pigment optical density (MPOD)~\cite{berendschot2006macular}, is

\begin{equation}
    D(r) = A_1 10^{-\rho_1 r} + A_2 10^{-\rho_2 (r - \alpha_2)^2},
    \label{eqn:mpod}
\end{equation}

where these parameters have been measured to be $A_1 = 0.31\pm0.12$, $A_2 = 0.11\pm0.08$, $\rho_1 = 0.32^\circ\pm0.24^\circ$, $\rho_2 = 1.2(^\circ)^{-2} \pm 1.1(^\circ)^{-2}$, and $\alpha_2 = 0.70^\circ \pm 0.66^\circ$~\cite{berendschot2006macular}.
A state with OAM value $\ell = 13$ was chosen because this state produced a significant number of azimuthal fringes without reaching the spatial resolution limit of our spatial light modulator (SLM).
The structured state generated by the SLM is a polarization-coupled OAM state described by,

\begin{equation}
    \frac{1}{\sqrt{2}}(1-\Pi[r/c])(e^{i (13 \phi + 2\theta t)}\ket{R} + \ket{L}),
    \label{eqn:spin-orbit-state}
\end{equation}

where we have used bra-ket notation for the right, $\ket{R}=\big(\begin{smallmatrix} 1 \\ 0 \end{smallmatrix}\big)$, and left, $\ket{L}=\big(\begin{smallmatrix} 0 \\ 1 \end{smallmatrix}\big)$, circular polarization states, $(r, \phi)$ for the cylindrical coordinate system, $\Pi[r/c]$ is the unit pulse function that sets the size ($c$ in Fig.~\ref{fig:apparatus}(a)) of the central obstruction area, and $2\theta t$ is a time-varying phase-shift.
The polarization state after each element in the device is described in the Appendix. 
The macula can be modeled by a partially polarizing optical element of the form,

\begin{equation}
    \hat{U}_M = \frac{D(r)}{2}\begin{pmatrix}
    1 & e^{-i 2\phi} \\
    e^{i 2\phi} & 1
    \end{pmatrix} + (1-D(r))\mathbf{1},
    \label{eqn:macula_polarizer}
\end{equation}

where $D(r)$ is the MPOD, and $\mathbf{1}$ is the identity operator.
The combination of this polarizing filter in the macula with a structured light state will produce a number of azimuthally varying fringes, where the number of fringes is given by $N = |\ell - 2|$~\cite{sarenac2020direct}.
The time-varying phase-shift, $2\theta t$, causes these entoptic fringes to rotate, either clockwise or counter-clockwise. 

The entoptic pattern created by this structured state and the macula model in Equations~\ref{eqn:mpod},~\ref{eqn:spin-orbit-state} is shown in Figure~\ref{fig:apparatus}(a), using the mean parameter values measured by~\cite{berendschot2006macular}.
The human cornea has been shown to possess birefringence, oriented about a roughly horizontal axis~\cite{van1987corneal}.
In the special case of a uniformly polarized beam, $\ell=0$, the two azimuthal fringes of Haidinger's Brush are visible, but the rotation direction of the pattern is reversed for large corneal birefringence values.
With the $\ell = 13$ OAM state, the perceived rotation direction of the entoptic phenomenon is insensitive to corneal birefringence, see the Appendix for a detailed derivation.

It has been observed that the visibility of stationary entoptic patterns created by polarized light structures will reduce over time due to visual adaptation.
Therefore, any polarization pattern projected onto the retina must vary with time.
This can be accomplished using a rotating linear polarizer that will modify the phase, $\theta$ in Equation~\ref{eqn:spin-orbit-state}, and cause the polarization-coupled OAM state to rotate.
Using the ability to control the polarization structure projected on the retina, limited only by the resolution of the SLM, we can measure the apparent size of this entoptic phenomenon through a perception task.

A schematic of the device used in this experiment is shown in Figure~\ref{fig:apparatus}(b).
The polarization-coupled OAM-state is created with a $\SI{450}{\nano\meter}$ wavelength fiber-coupled diode laser, attenuated to $\SI{150}{\nano \watt}$ at the location of the pupil.
After exiting the coupler, the light is prepared in a circularly polarized state, followed by a rotating polarizer and quarter waveplate combination which allows for any polarization state orthogonal to the horizontal axis to be prepared at equal intensity.
This light is then reflected from an SLM, whose optical axis is aligned horizontally.
The SLM is placed at one output of a beamsplitter, while a red laser and pinhole, creating a fixation target light for participants, are placed at the other output port.
Both the SLM screen and pinhole are then imaged and expanded by a $4f$-imaging system.
A 20D Volk lens is then used to map the image created by the $4f$-imaging system directly onto a participant's retina.
This direct projection of the polarization-coupled OAM state onto the retina removes the effects of state propagation seen in~\cite{sarenac2020direct} and enables the use of a variable obstruction on the SLM screen.
Real-time in-vivo imaging is enabled by an 80R/20T plate beamsplitter and a linear polarizer, both of which can be moved into the beam via translation stages.
Light reflected off a participant's retina then passes through an analyzer, anti-aligned to the movable polarizer placed before the beamsplitter, which eliminates corneal reflections from the image.
Finally, the retina is imaged with the ZWO camera shown in Figure~\ref{fig:apparatus}(b).

The high refresh-rate SLM used to generate this state allows for the real-time control of an arbitrary polarization pattern that illuminates a participant's retina directly.
Therefore, we can obstruct the central region of a structured light state to determine the spatial extent of the entoptic pattern generated by said profile, as can be seen in Figure~\ref{fig:apparatus}(a).
To ensure that participants are not able to distinguish the motion of the entoptic pattern from any other motion, the rotating polarizer used to generate rapid motion in the polarization profile is set to always rotate in the same direction.
Distinct motion in the entoptic stimulus can be generated by changing the orientation of the quarter waveplate following the rotating polarizer.
Note that this quarter waveplate remains stationary during a trial.

\section{Empirical Experiment}
Twenty-three participants were recruited to perform the discrimination task.
All participants provided informed consent and were treated in accordance with the Declaration of Helsinki.
All research procedures received approval from the University of Waterloo Office of Research Ethics.

The retina of each participant was illuminated with a structured light state, where an OAM number of $\ell=13$ was prepared on the right-circular polarization state, resulting in $N = 11$ fringes when filtered with the polarizer described in Equation~\ref{eqn:macula_polarizer}.
Aligned to the center of this OAM state was a $\SI{50}{\micro\meter}$ pinhole illuminated by a red laser, creating a $\approx 1^\circ$ visual angle guide light for participants to fixate on.
Each trial consisted of $\SI{500}{\milli\second}$ of an OAM state, which produced 11 azimuthal fringes, rotating either clockwise (CW) or counterclockwise (CCW).
After the $\SI{500}{\milli\second}$ presentation, the structured light pattern was blocked by a shutter, while the fixation target remained on, and the participant indicated the direction of pattern rotation.
Each new trial began with the presentation of a stationary structured light state pattern to allow the participant to verify visibility, and the onset of the task rotation was participant-controlled.

In order to measure the eccentric extent of polarization sensitivity, a circular obstruction was centered on the fixation point, and the radius was varied following the 2-up/1-down staircase method~\cite{levitt1971transformed}.
Two consecutive correct answers resulted in an increase in the radius of the circular obstruction placed at the center of the structured light state, while one incorrect response resulted in a decrease in the radius of the central obstruction.
This procedure enables the calculation of an obstruction threshold corresponding to 70.7\% performance accuracy, where a larger threshold indicates a larger eccentric range of polarization sensitivity.

\begin{figure}
    \centering
    \includegraphics[width=0.45\textwidth]{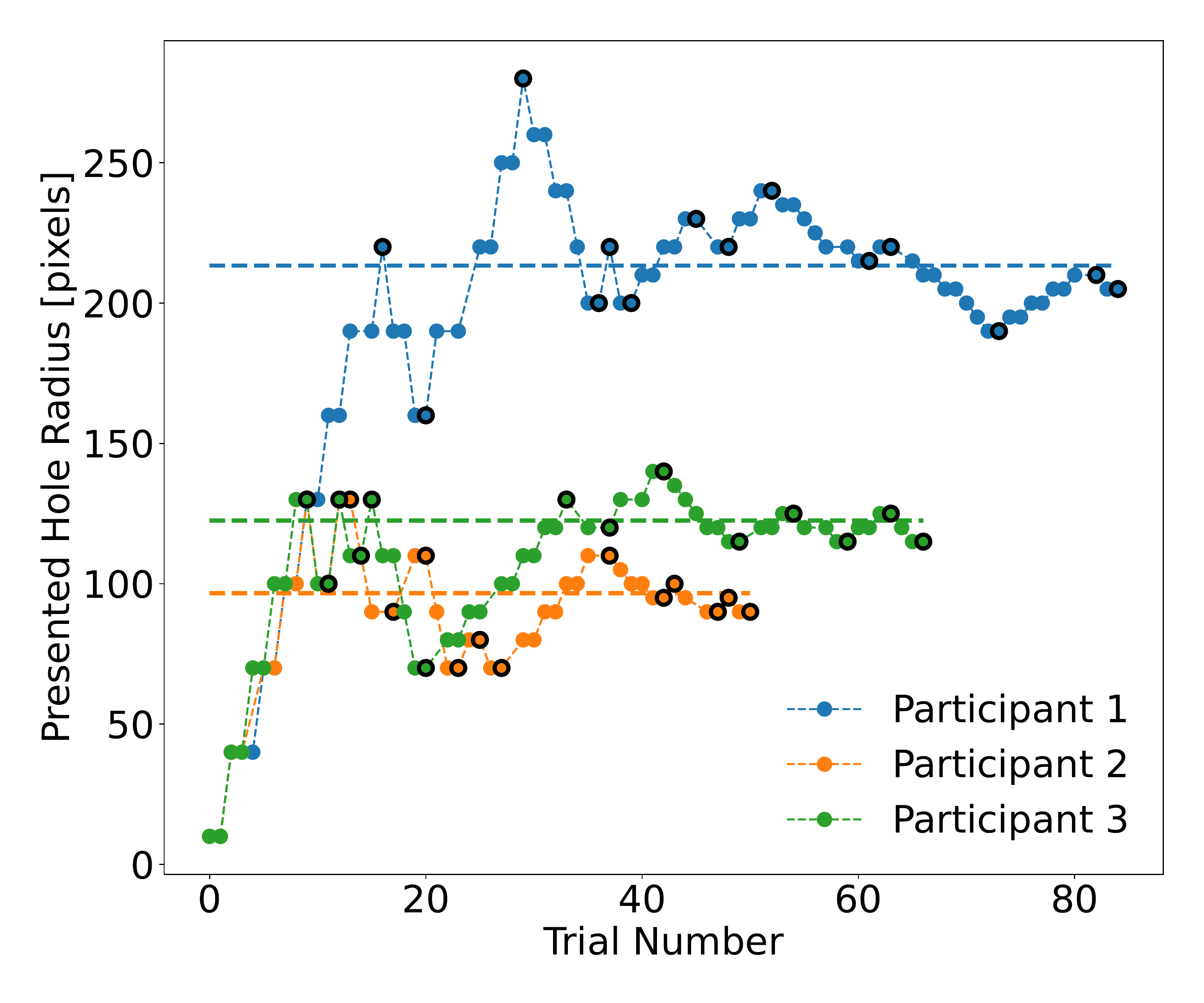}
    \caption{Select examples of participant staircase results.
    The final six reversal points, labeled as black dots, are averaged to determine the radius at which the structured light stimulus is perceivable at 70.7\% correct.
    Three example staircases are shown.
    }
    \label{fig:example_staircase}
\end{figure}

The staircase terminated after 14 reversals (that is, a change in hole-size progression from increasing to decreasing or vice versa) or after 90 total trials.
The initial obstruction was 10 pixels in radius, equal to the radius of the red fixation point.
The initial step size of the change in central obstruction visual angle was $2.70^\circ$, became $1.80^\circ$ after three reversals, then became $0.90^\circ$ after six reversals, and finally became $0.45^\circ$ after nine reversals.
Each person's eccentric obstruction threshold is calculated by taking the arithmetic mean of the final six reversal points.
If the participant completed 90 trials, then the final point was treated as a reversal point.
Example staircases and thresholds are shown in Figure~\ref{fig:example_staircase}.
Before performing the thresholding task, all participants performed an initial familiarization task.
The initial obstruction size ($0.90^\circ$) was presented for several seconds 10 times, and all participants achieved at least 70$\%$ discrimination accuracy.
Participants whose staircase results show more than 3 reversal points at the minimum obstruction radius are considered to have ``failed'' the perception task.
Altogether, 18 out of 24 total participants successfully completed the task with a valid threshold.
While these ``failed'' participants correctly perceived the entoptic phenomenon rotation direction when presented for no less than 2 seconds, the stimulus presentation time of $\SI{500}{\milli\second}$ inhibited these participants' ability to discern a rotation direction.
One participant produced a threshold visual angle that was 3.3 standard deviations beyond the mean value and was removed on the assumption of improper fixation.

\begin{figure}
    \centering
    \includegraphics[width=0.45\textwidth]{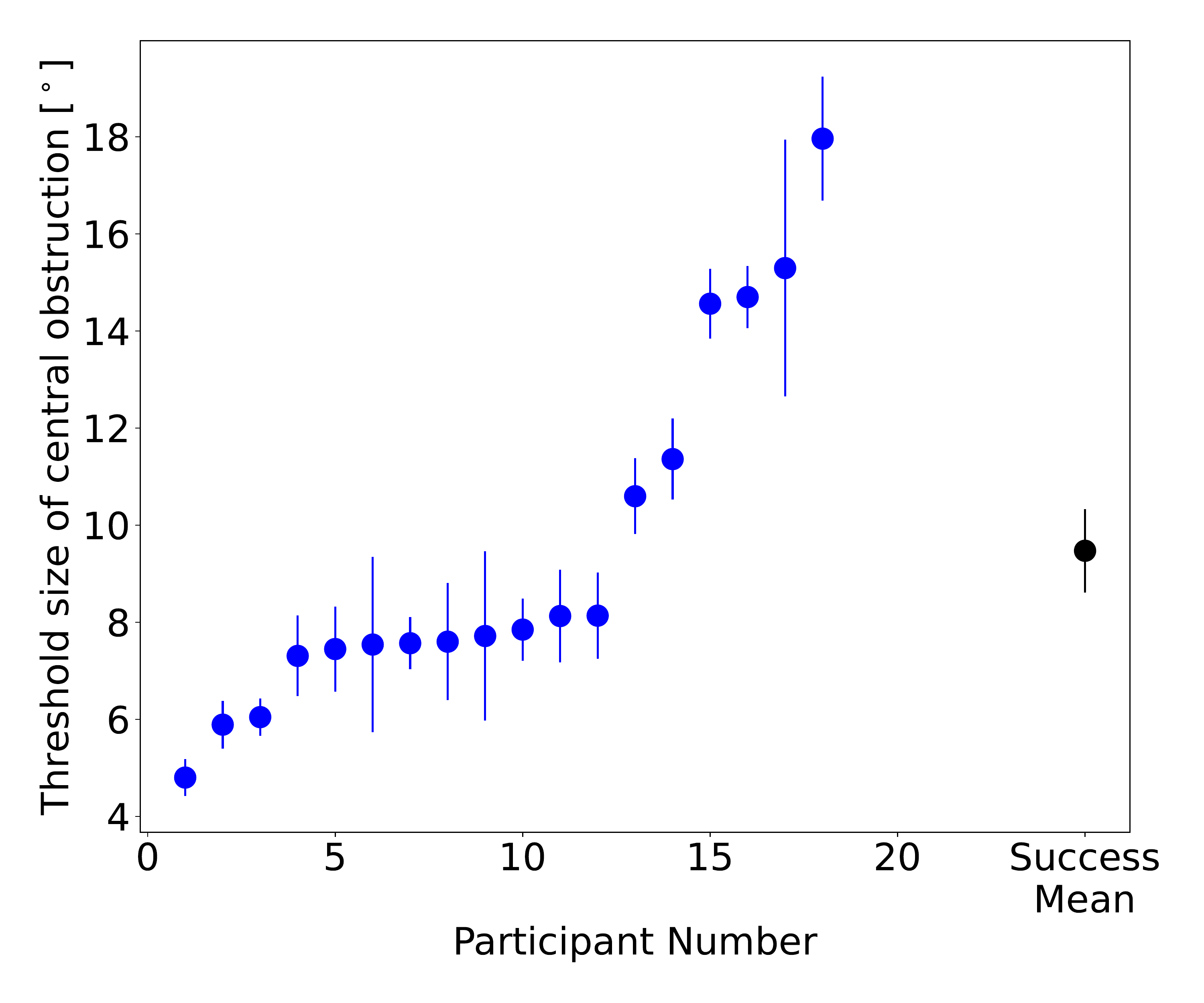}
    \caption{Threshold results from all participants who successfully complete the perception task, shown as blue points.
    The threshold value was measured by averaging the final six reversal points from each participant's staircase, with the standard deviation of the reversal points indicated as error bars.
    The threshold was converted to degrees of visual angle impinging on the retina by coregistering a structured light image of the retina with a commercial fundus image and calculating the size of landmarks.
    The average threshold visual angle was $9.5^\circ \pm 0.9^\circ$.
    }
    \label{fig:staircase_results}
\end{figure}

\section{Results}
\begin{figure*}
    \centering
    \includegraphics[width=0.9\textwidth]{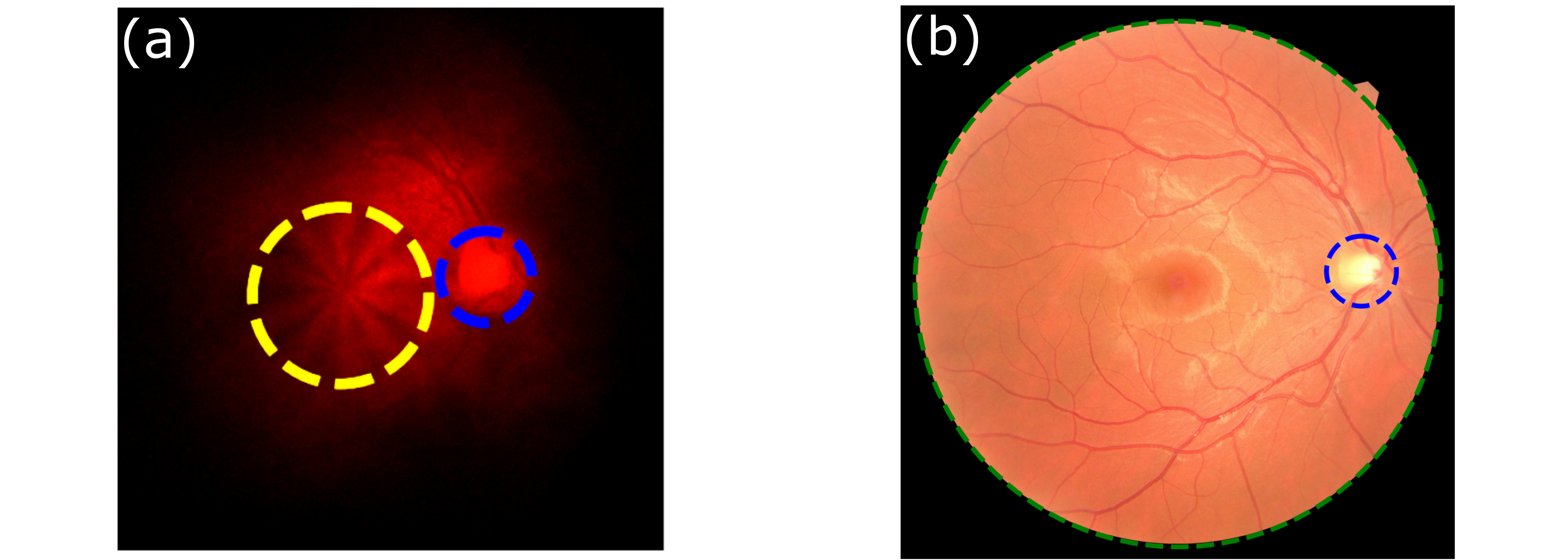}
    \caption{Two example retinal images from a single participant.
    (a) Retinal image taken with the structured light imaging device.
    The incident light is passed through a polarizer in order to display the radial pattern of the structured light state.
    By displaying structured light states limited to a set radius, we can measure the size of these patterns relative to features on each participant's retina.
    (b) Retinal image captured by a Nidek MP-3 fundus camera.
    The size of features in the optic nerve head can be used to relate the size of various structured light patterns.
    These two images allow us to verify the detection threshold radius in degrees of visual angle, accounting for differences on an individual level.}
    \label{fig:retinal_imaging}
\end{figure*}

From each participant's perception threshold in SLM pixel units, it is possible to calculate the threshold in degrees of visual angle.
To achieve this, retinal images are taken using the described structured light device to determine the spatial extent of the central field obstruction on the retina through an image that shows both the optic disk and the structured light pattern.
As a confirmation, retinal images are also taken using a Nidek MP-3 fundus camera and the size of the optic disk in degrees of visual angle can be compared between these two photos.
An example of the image captured by the structured light imaging device can be seen in Figure~\ref{fig:retinal_imaging}(a) and the corresponding image produced by the commercial retinal camera can be seen in  Figure~\ref{fig:retinal_imaging}(b).

The results for each participant, converted from pixel units into degrees of vision, are shown in Figure~\ref{fig:staircase_results}, where the mean value is determined by the average of the final six reversal points and the error bars are given by the standard deviation of these reversal points.
As can be seen in Figure~\ref{fig:staircase_results}, the average threshold was $9.5^\circ \pm 0.9^\circ$ with a 95\% confidence interval of [$5.83^\circ$, $13.12^\circ$].
These results show that all participants observe an entoptic pattern that is larger than the mean size of Haidinger's Brush measured in~\cite{coren1971use}.

\section{Discussion}
In this work, we have directly mapped a polarization-coupled OAM state onto the human retina.
This negates propagation effects seen in~\cite{sarenac2020direct}, enabling the use of a variable obstruction on the SLM screen itself.
Using this variable obstruction, we have directly measured the extent of the entoptic phenomenon created by a polarization-coupled OAM state.
This, combined with structured light retinal imaging has allowed for the measurement of these entoptic phenomena in degrees of visual angle.
Between participants, there is significant variation in the visual angle this entoptic pattern subtends.
Patient history, refractive error, fundus images, and optical coherence tomography data were gathered on each patient during this study.
No significant correlation is observed between perceived entoptic pattern size and refractive error, including spherical, cylindrical, and spherical equivalent refractive errors.
One possible source of the variation between patient thresholds is individual differences in macular pigment density.

The resulting mean apparent size of the entoptic pattern with $11$ azimuthal fringes, $9.5^\circ \pm 0.9^\circ$, exceeds measurements of the apparent size of Haidinger's Brush, $3.75^\circ$~\cite{coren1971use}.
The diagnostic potential of Haidinger's Brush perception tasks has previously been established~\cite{rothmayer2007nonlinearity, muller2016perception}.
The significant increase in the apparent size and clarity of entoptic phenomenon produced by polarization-coupled OAM stimuli suggests a greater utility of structured light stimuli as a perception test to evaluate macular health.
These advancements open new avenues of research whereby the extent of patterns created by structured light stimuli can be used to measure properties of the macula and spatially map changes in the macula through polarized light perception.
Future studies will relate direct measurements of spatially-dependent changes in polarization perception to AMD and other retinal disorders.

\begin{acknowledgments}
This work was supported by the Canadian Excellence Research Chairs (CERC) program, the Natural Sciences and Engineering Research Council of Canada (NSERC) Discovery program, the Collaborative Research and Training Experience (CREATE) program, and the Canada First Research Excellence Fund (CFREF).
This work was also supported by the Government of Canada’s New Frontiers in Research Fund (NFRF) NFRFE-2019-00446, the Hong Kong SAR government, and the InnoHK Centre for Eye and Vision Research (CEVR).
\end{acknowledgments}

\bibliography{main}

\clearpage
\section*{Appendix}
\label{sec:appendix}
\subsection{Coordinate System}

Here we quantize the polarization state along the circular polarization axis.
The real-space coordinate system is defined with $z$ following beam propagation, $x$ on the horizontal, and $y$ on the vertical with positive $y$ pointing `up'.
In this system, the diagonal state $\ket{D}$ is pointed in the positive $x$ and $y$ quadrant.
Up to a normalization factor, these states are given by,

\begin{align*}
    \ket{R} = \big(\begin{smallmatrix} 1 \\ 0 \end{smallmatrix}\big),& \ket{L} = \big(\begin{smallmatrix} 0 \\ 1 \end{smallmatrix}\big), \\ 
    \ket{H} = \big(\begin{smallmatrix} 1 \\ 1 \end{smallmatrix}\big),& \ket{V} = \big(\begin{smallmatrix} 1 \\ -1 \end{smallmatrix}\big), \\ 
    \ket{D} = \big(\begin{smallmatrix} 1 \\ i \end{smallmatrix}\big),& \ket{A} = \big(\begin{smallmatrix} 1 \\ -i \end{smallmatrix}\big).\\ 
\end{align*}

\subsection{Polarization State in the Apparatus}

Incident to the rotating polarizer, the laser is prepared in $\ket{\psi_0} = \ket{R}$ state.
Following the rotating polarizer, the state is described by,

\begin{equation*}
    \ket{\psi_1} = \ket{R} + e^{i 2\theta}\ket{L},
\end{equation*}

where the angle $\theta$ is defined relative to the horizontal axis in real space.
Following the rotating polarizer, a rotating quarter-waveplate is set to either the diagonal or anti-diagonal axis.
This produces one of two states,

\begin{align*}
    \ket{\psi_{2, D}} &= \ket{H} + e^{i 2\theta}\ket{V}, \\
     &= cos(\theta)\ket{R} - sin(\theta)\ket{L}, \\
    \ket{\psi_{2, A}} &= \ket{V} + e^{i 2\theta}\ket{H}, \\
     &= cos(\theta)\ket{R} + sin(\theta)\ket{L},
\end{align*}

where the waveplate setting has been denoted with the subscript $D$ or $A$.
The spatial light modulator will then write a spatially dependent phase on the horizontal polarization state, for example, we will choose an orbital angular momentum (OAM) state with OAM number $\ell$,

\begin{align*}
    \ket{\psi_{3, D}} &= e^{-i \ell \phi}\ket{H} + e^{i 2\theta}\ket{V}, \\
    \ket{\psi_{3, A}} &= e^{-i \ell \phi}e^{i 2\theta}\ket{H} + \ket{V}.
\end{align*}

The spatial light modulator is followed by a fixed quarter waveplate which is set to diagonal, mapping the states $\ket{H} \mapsto \ket{L}$ and $\ket{V} \mapsto \ket{R}$.
This produces one of two states,

\begin{align*}
    \ket{\psi_{4, D}} &= e^{i 2\theta}\ket{R} + e^{-i \ell \phi}\ket{L}, \\
    \ket{\psi_{4, A}} &= \ket{R} + e^{-i \ell \phi}e^{i 2\theta}\ket{L}.
\end{align*}

\subsection{Entoptic Pattern, Without Corneal Birefringence}\label{app:entoptic}

The pattern perceived by a test subject can be modeled using a radial polarization filter, given by Equation~\ref{eqn:macula_polarizer}.
Assuming the macula perfectly polarizes, ie. $D(r) = 1$, this filter produces an intensity pattern, given by,

\begin{align*}
    I_D &= \frac{1}{4}\big(\begin{smallmatrix} e^{-i 2\theta} & e^{i\ell\phi} \end{smallmatrix}\big) \big(\begin{smallmatrix} 1 & e^{-i2\theta} \\ e^{i2\theta} & 1 \end{smallmatrix}\big) \big(\begin{smallmatrix} e^{i 2\theta} \\ e^{-i\ell\phi} \end{smallmatrix}\big), \\
    I_D &= \frac{1}{4}(2 + e^{-i 2\theta} e^{-i (\ell+2)\phi} + e^{i 2\theta} e^{i (\ell+2)\phi}), \\ 
    &= \cos^2 \left(\frac{\ell+2}{2}\phi + \theta \right), \\~\nonumber\\
    I_A &= \frac{1}{4}\big(\begin{smallmatrix} 1 & e^{-i 2\theta}e^{i\ell\phi} \end{smallmatrix}\big) \big(\begin{smallmatrix} 1 & e^{-i2\theta} \\ e^{i2\theta} & 1 \end{smallmatrix}\big) \big(\begin{smallmatrix} 1 \\ e^{-i\ell\phi}e^{i 2\theta} \end{smallmatrix}\big), \\
    I_A &= \frac{1}{4}(2 + e^{i 2\theta} e^{-i (\ell+2)\phi} + e^{-i 2\theta} e^{i (\ell+2)\phi}), \\
    &= \cos^2 \left(\frac{\ell+2}{2}\phi - \theta \right).
\end{align*}

The above shows that, for any $\ell$ value, this state produces azimuthal fringes which rotate clockwise or counter-clockwise depending on the rotation waveplate setting.
The effect of corneal birefringence can be considered by adding another waveplate element before the macular polarizer.
Consider the strongest possible corneal effect, namely a half-waveplate oriented horizontally.
The effect of this element is to map the states $\ket{R} \mapsto \ket{L}$ and $\ket{L} \mapsto \ket{R}$.

\begin{align*}
    I'_D &= \frac{1}{4}\big(\begin{smallmatrix} e^{i\ell\phi} & e^{-i 2\theta} \end{smallmatrix}\big) \big(\begin{smallmatrix} 1 & e^{-i2\theta} \\ e^{i2\theta} & 1 \end{smallmatrix}\big) \big(\begin{smallmatrix} e^{-i\ell\phi} \\ e^{i 2\theta} \end{smallmatrix}\big), \\
    I'_D &= \frac{1}{4}(2 + e^{i 2\theta} e^{i (\ell-2)\phi} + e^{-i 2\theta} e^{-i (\ell+2)\phi}), \\
    &= \sin^2 \left(\frac{\ell-2}{2}\phi + \theta \right), \\~\nonumber\\
    I'_A &= \frac{1}{4}\big(\begin{smallmatrix} e^{i 2\theta}e^{i\ell\phi} & 1 \end{smallmatrix}\big) \big(\begin{smallmatrix} 1 & e^{-i2\theta} \\ e^{-i2\theta} & 1 \end{smallmatrix}\big) \big(\begin{smallmatrix} e^{-i\ell\phi}e^{i 2\theta} \\ 1 \end{smallmatrix}\big), \\
    I'_A &= \frac{1}{4}(2 + e^{-i 2\theta} e^{i (\ell-2)\phi} + e^{i 2\theta} e^{-i (\ell+2)\phi}), \\ 
    &= \sin^2 \left(\frac{\ell-2}{2}\phi - \theta \right).
\end{align*}

Comparing these two cases with $\ell = 5$, a participant with a corneal birefringence of $\gamma_{cornea} = 180^\circ$ will see $3$ azimuthal fringes while a participant with no corneal birefringence will see $5$ lines.
Critically, both participants will observe rotation clockwise or counter-clockwise at the same waveplate setting.
In the case of a uniform polarization state, producing a Haidinger's brush entoptic pattern, the participants will observe,

\begin{align*}
    I_D = \cos^2 \left( \phi + \theta \right), ~& I_A = \cos^2 \left( \phi - \theta \right), \\
    I'_D = \cos^2 \left( \phi - \theta \right), ~& I'_A = \cos^2 \left( \phi + \theta \right),
\end{align*}

thus for the same waveplate setting, each participant will observe an opposite rotation direction.

\subsection{Entoptic Pattern, With Corneal Birefringence}

Assuming a horizontal orientation for the birefringence inherent in the cornea, we can generalize the above result for any rotation about the horizontal polarization by an angle, $\beta$.
The effect of the cornea is given by the operation,

\begin{equation*}
    \hat{U}_C = \begin{bmatrix}
\cos\left( \frac{\beta}{2} \right) & -i \sin\left( \frac{\beta}{2} \right) \\
\sin\left( \frac{\beta}{2} \right) & \cos\left( \frac{\beta}{2} \right).
\end{bmatrix}
\end{equation*}

The state after the cornea is given by,

\begin{align*}
    \ket{\psi_{5,D}} &= \hat{U}_C\ket{\psi_{4, D}}, \\
    &= \left( e^{i 2 \theta}\cos \left( \beta/2 \right) -ie^{-i \ell \phi}\sin \left( \beta/2 \right) \right)\ket{R} \\ 
    &+ \left( ie^{i 2 \theta}\sin \left( \beta/2 \right) -e^{-i \ell \phi}\cos \left( \beta/2 \right) \right)\ket{L}, \\
    \ket{\psi_{5,A}} &= \hat{U}_C\ket{\psi_{4, A}}, \\
    &= \left( \cos \left( \beta/2 \right) -ie^{i 2 \theta}e^{-i \ell \phi}\sin \left( \beta/2 \right) \right)\ket{R} \\
    &+ \left( i\sin \left( \beta/2 \right) + e^{i 2 \theta}e^{-i \ell \phi}\cos \left( \beta/2 \right) \right)\ket{L}.
\end{align*}

We can calculate the intensity profile for this generalized case using this state $\ket{\psi_{5,D}}$ or $\ket{\psi_{5,A}}$,

\begin{align*}
    I_D &= \left| \bra{M}\ket{\psi_{5,D}} \right|^2 \\
    &= 1/2 + 1/2\cos([\ell+2]\phi+2\theta)\cos^2(\beta/2) \\ 
    &\hspace{1cm}- 1/2\sin(\beta/2)\cos(\beta/2)\sin(\ell\phi + 2\theta) \\
    &\hspace{1cm}- 1/2\sin(\beta/2)\cos(\beta/2)\sin(2\phi) \\
    &\hspace{1cm}- 1/2\cos([\ell-2]\phi+2\theta)\sin^2(\beta/2), \\~\nonumber\\
    I_A &= \left| \bra{M}\ket{\psi_{5,A}} \right|^2 \\
    &= 1/2 + 1/2\cos([\ell+2]\phi-2\theta)\cos^2(\beta/2) \\ 
    &\hspace{1cm}- 1/2\sin(\beta/2)\cos(\beta/2)\sin(\ell\phi - 2\theta) \\
    &\hspace{1cm}- 1/2\sin(\beta/2)\cos(\beta/2)\sin(2\phi) \\
    &\hspace{1cm}- 1/2\cos([\ell-2]\phi-2\theta)\sin^2(\beta/2).
\end{align*}

While these patterns do not have a well-defined number of equal contrast fringes, the phenomenon seen in the previous case holds.
That is, for OAM states with $\ell \neq 0$, the entoptic pattern caused by the macula appears to rotate clockwise/counter-clockwise for every participant regardless of their corneal birefringence value.
For the case of $\ell = 0$, there exists a second birefringence value of interest, $\beta = 90^\circ$, where no continuous motion of Haidinger's brush is apparent.

\end{document}